# Performance Analysis of Spin Transfer Torque Random Access Memory with cross shaped free layer using Heusler Alloys by using micromagnetic studies

Tangudu Bharat Kumar[1], Bhaskar Awadhiya[1], E.MeherAbhinav[1] and Bahniman Ghosh[2,3] and Bhupesh Bishnoi[2]

[1]Department of Electronics and Communications, Manipal Institute of Technology, Manipal University, Manipal, India.

[2]Department of Electrical Engineering, Indian Institute of Technology Kanpur, Kanpur 208016, India,Tel: +91 8960436590, Fax: +91-512-2590063

[3]Microelectronics Research Center, 10100 Burnet Road, Bldg. 160, University of Texas at Austin, Austin, TX, 78758, USA.

Abstract:

We investigated the performance of spin transfer torque random access memory (STT-RAM) cell with cross shaped Heusler compound based free layer using micromagnetic simulations. We designed the free layer using Cobalt based Heusler compounds. Here in this paper, simulation results predict that switching time from one state to other state is reduced. Also it is examined that critical switching current density to switch the magnetization of free layer of STT RAM cell is reduced.

Introduction:

Spin transfer torque (STT) effect is one of the recent technologies to achieve the non-volatility, high speed, reduction in power consumption, large storage density in memory devices [1-8]. Also STT based memory devices provide better performance with respect to scalability of dimensions of a device. The simplest STTRAM cell is a combination of magnetic tunneling junction (MTJ) and access transistor. Free layer (FL) and Pinned layer (PL), separated by non magnetic insulating layer are together called MTJ. FL and PL are made of ferromagnetic materials. In these devices, data is stored in terms of magnetization direction of free layer. To switch the magnetization direction of the free layer, large current densities are required which limit the storage density of memory device. Some groups individually proposed the methodologies to store the multiple bits by connecting the MTJ devices in series [9] or creating domains in free layer [10] or using stacked MTJs [11].

Recently Roy et al. designed MTJ with cross shaped free layer to store the two bits in a single MTJ which avoid the problem occurring due to increasing storage density [12]. They used Cobalt as a ferromagnetic material to design the free layer.

Critical switching current density can be modeled by considering the finite penetrating depth of spin current and spin pumping [13-15]. In macro spin model, critical switching current density ($J_{c0}$) at Zero temperature can be defined as,

$$J_{co} = \frac{2e\alpha M_s t(H + H_k + 2\pi M_s)}{\hbar P} \quad (1)$$

Here, $M_s$ is saturation magnetization, t is free layer thickness, α is damping constant, $H_k$ is anisotropy field, H is external field, P is spin polarization factor and $\hbar$ is reduced Planck constant.

By looking at Eq. (1), to reduce the switching current density, we notice that we need to choose the material which has low $M_s$, low α and high P. Heusler compounds satisfy the above requirements [16-17, 19-22]. Ideally heusler compounds have 100% spin polarization at the Fermi level [16]. Cobalt based heusler compounds have the advantage of low mismatch of lattice constant, high spin polarization factor and high tunneling magneto resistance (TMR) ratio [23-24]. $Co_2MnSi$ (CMS) produces high TMR values at low temperatures and TMR value of CMS decreases with increasing temperature. $Co_2FeAl_{0.5}Si_{0.5}$ (CFAS) also produces high TMR values for all temperature and TMR value of CFAS is weakly dependent of temperature [18].

In our paper, we select two Cobalt based heusler compounds which have low α, low $M_s$, high TMR values and high spin polarization factor. We choose $Co_2MnSi$ (CMS) and $Co_2FeAl_{0.5}Si_{0.5}$ (CFAS) to design free layer mentioned by Roy et al., and we simulate the structure using micromagnetic simulator called Object Oriented Micromagnetic Framework (OOMMF) [27]. We show that switching time and critical switching current density are less when we design the free layer with heusler compounds.

## Theoretical Model:

We take Ref. [12] as the reference model with dimensions of cross shape to be 100×140×2 $nm^3$. Aspect ratios of shorter arm are $l_1/w=2$, $l_2/w=3$, respectively. The stable magnetization states are shown in fig.1 of Ref. [12]. The material parameters for Cobalt are taken as follows [25], saturation magnetization ($M_s$) is 1400×$10^3$ A/m, uniform exchange constant (A) is 30×$10^{-12}$ J/m, spin polarization constant(P) is 0.4, damping constant (α) is 0.01. For CMS [26] the values of the respective parameters are as follows, $M_s$ is 800×$10^3$ A/m, A is 2.35×$10^{-12}$ J/m, P is 0.56 and α is 0.008. Also, for CFAS [28] the values are, $M_s$ is 900×$10^3$ A/m, A is 2.0×$10^{-11}$ J/m, P is 0.76 and α is 0.01. Since we are considering the shape anisotropy only, we are not using any crystalline anisotropy. Magnetization dynamics can be solved by Landau-Lifshitz-Gilbert equation with an added STT term,

$$\frac{d\vec{m}}{dt} = -|\gamma|\vec{m} \times \vec{H}_{eff} + \alpha\left(\vec{m} \times \frac{d\vec{m}}{dt}\right) + |\gamma|\beta\epsilon(\vec{m} \times \vec{m}_p \times \vec{m}) - |\gamma|\beta\epsilon'(\vec{m} \times \vec{m}_p) \quad (2)$$

Where $\vec{m}$ is reduced magnetization, α is damping constant, γ is gyromagnetic ratio, $\beta = \left|\frac{\hbar}{\mu_\circ e}\right|\frac{J}{tM_s}$, $M_s$ is saturation magnetization in A/m, t is thickness of free layer in meters, J is charge current density in A/m$^2$, $\vec{m}_p$ is unit polarization direction of the spin polarized current, $\epsilon = \frac{P\Lambda^2}{(\Lambda^2+1)+(\Lambda^2-1)(\vec{m}\cdot\vec{m}_p)}$, P is spin polarization constant and ϵ' is secondary spin transfer term. In Eq.2, spin polarization dependence of $\vec{m}$ can be explained by ϵ. As TMR is dependent on cosine of angle between free layer and fixed layer magnetization, the authors proposed 22.5° as an angle between the short arm of free layer and fixed layer. Also ϵ' and $m_p$ are taken as 0.06 and (0.92 0.382 0) respectively. Initial magnetization can be taken in one of the four quadrants of xy-plane. For our calculations, we are taking Λ equal to 1.

## Results and Discussion:

We performed several micromagnetic simulations for a given cross shaped free layer structure using OOMMF with different types of materials (CMS, CFAS and Cobalt) by supplying negatively polarized current density (J) values ranging from 0 to 4×10$^{12}$ A/m$^2$. By the definition of critical switching current density, it is the value at which direction of magnetization changes to 180° by crossing energy barrier. In the absence of applied current density, free layer does not change its magnetization direction. Depending on the value of current density, short arm or long arm or both arms of the free layer change their direction of magnetization. If the J value is less, and if it is enough to cross the energy barrier, short arm of the free layer changes its magnetization ($m_x$) direction. And if J value is large enough to change the direction of magnetization of long arm ($m_y$), then both arms of the free layer change their directions. To change the direction of only long arm magnetization direction, we have to apply the two opposite current pulses successively.

We have performed this same procedure to the same structure but made with different types of materials Cobalt, CMS and CFAS by taking their material parameters into simulation. It is observed that in Fig.1 and Fig.2, switching time from state1 to state2 (fig.1 in Ref. [12]) is less compared to switching time for the same if the Cobalt made free layer is simulated. Also, it is observed that less switching time can be achieved for fewer values of current densities when structure is simulated with CMS and CFAS material parameters. After stopping the supply of current, system would reach the nearest stable state. It can be seen in Fig.1 and Fig. 2 at 6ηs and 1.5ηs respectively.

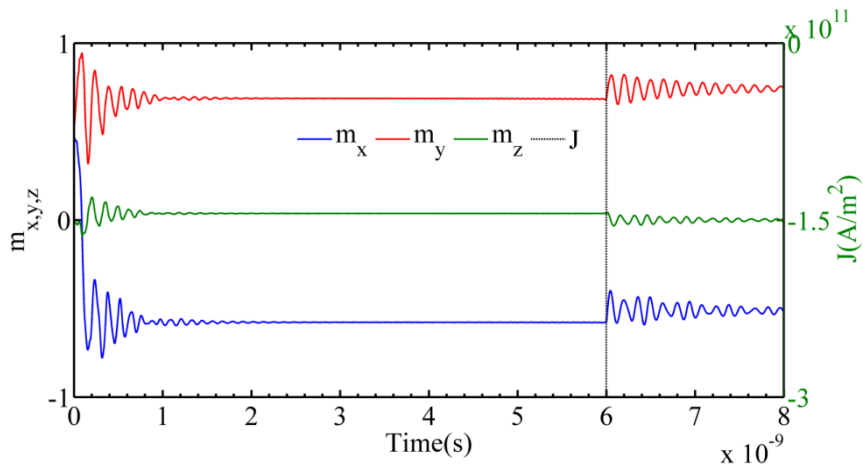

Fig.1. CFAS based free layer switching from state1 to state2 at J = -3×10$^{11}$ A/m$^2$.

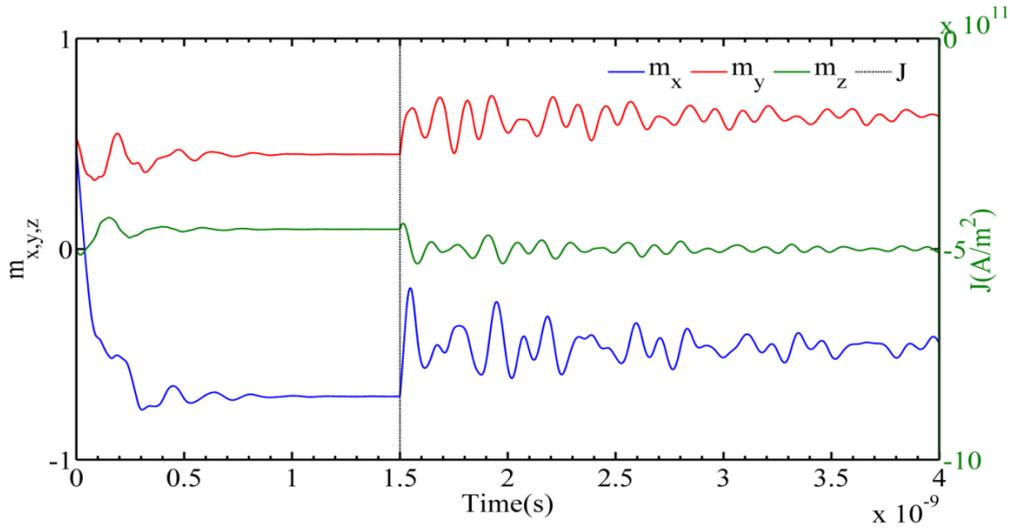

Fig.2. CMS based free layer switching from state1 to state2 at J = -1×10$^{12}$ A/m$^2$.

System will take some time to switch from one state to another state which we call as switching time (t$_s$). Variation of switching time from state1 to state2 with different current densities for three materials can be seen in Fig.3. It is seen from Fig.3 that CFAS free layer has less switching time than CMS and Cobalt based free layers at any particular current density value, ranging up to -4.5×10$^{11}$A/m$^2$.

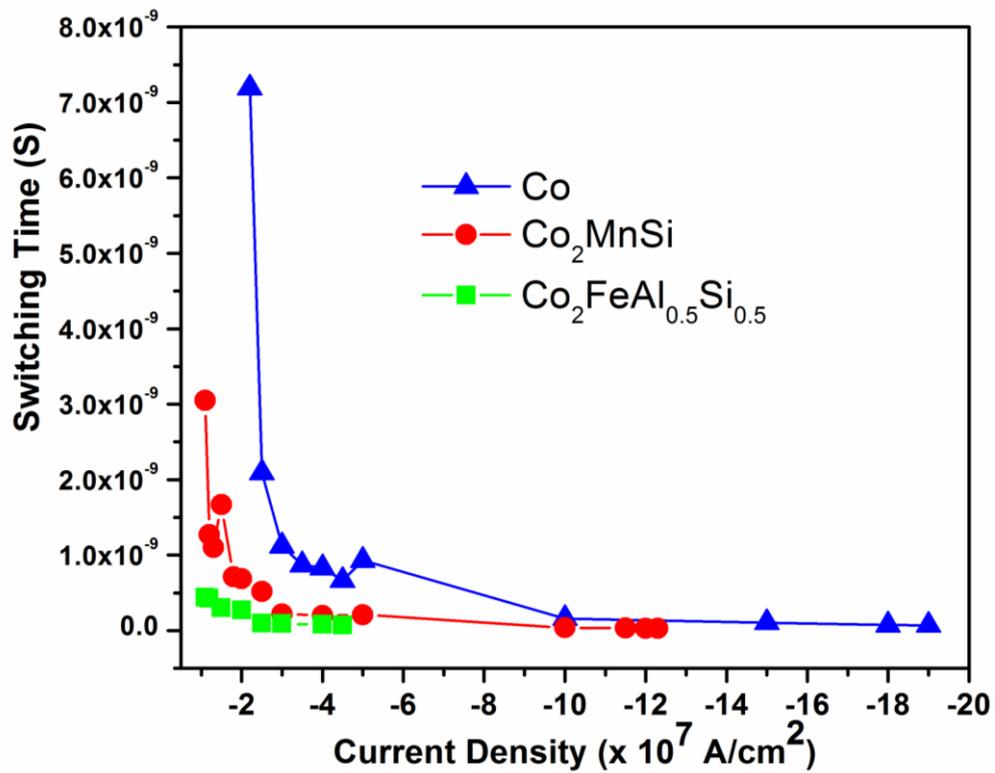

Fig.3. Variation of switching time from state1 to state2 with different current densities

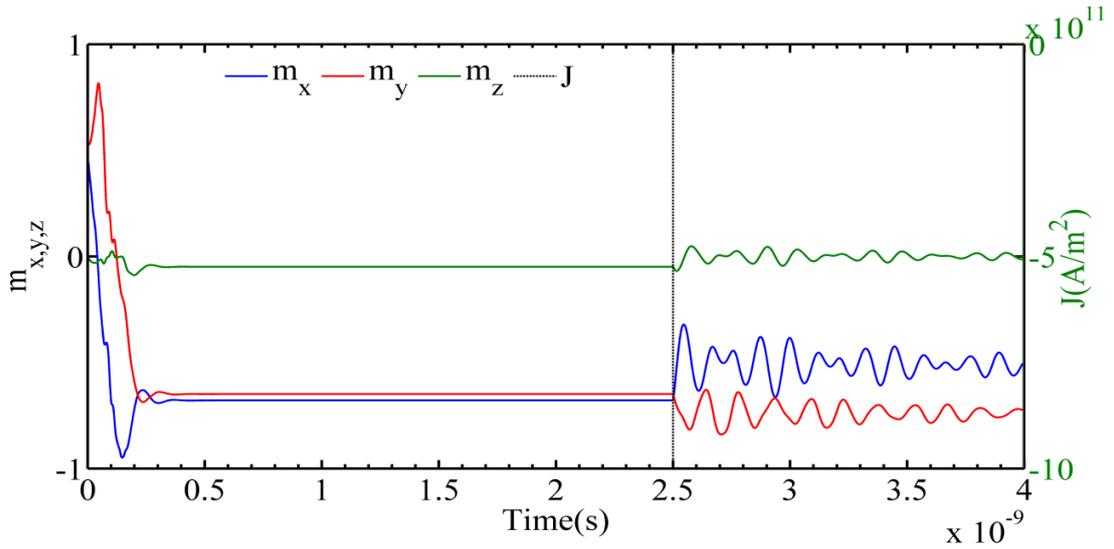

Fig.4. CFAS based free layer switching from state1 to state3 at $J = -1 \times 10^{12}$ A/m$^2$

Switching from first state to third state at $J=-1\times10^{12}$ A/m$^2$ can be observed in Fig.4. It is also observed from Fig.2 and Fig.4 that less amount of current density is required to switch from first state to third state for CFAS when compared to CMS. For cobalt based free layer, it required to have more amount of current density values to switch from first state to third state. Similarly switching from first state to third state for CMS based free layer at $J = -\times10^{12}$ A/m$^2$ can be shown in Fig.5.

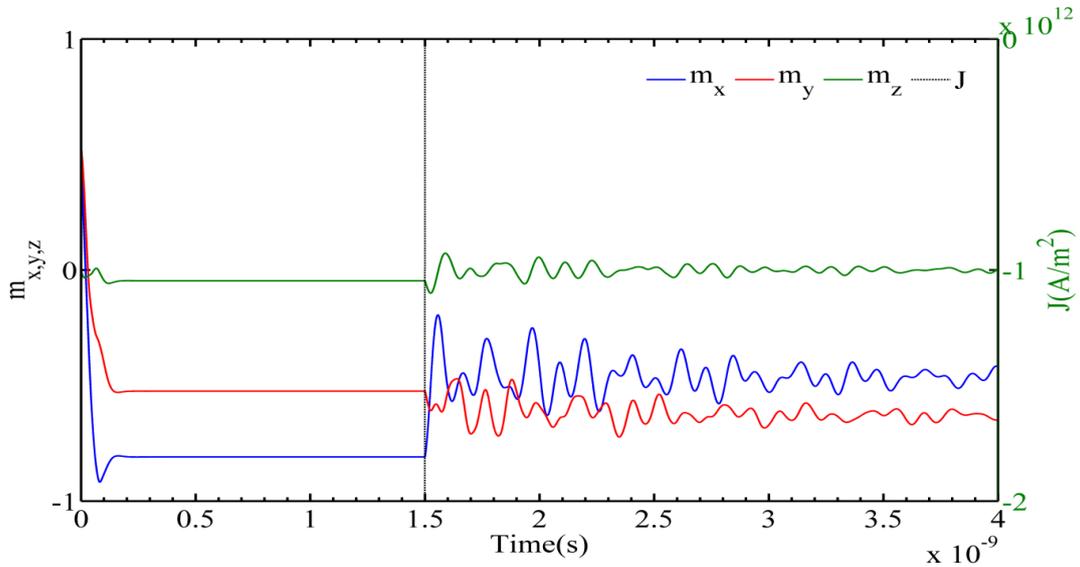

Fig.5. CMS based free layer switching from state1 to state3 at $J = -2\times10^{12}$ A/m$^2$

Fig.6 gives the brief explanation of the above discussion. It tells the different switching regions from first state to second state and third states. "Black" color bar refers to state1; "dark grey" bar refers to state2 region and "light grey" bar refers to state3 region. It can be seen that CFAS based free layer reaches first and third states for lesser values of current densities only. CMS based free layer switches to other states from first state for fewer values of current densities compared to cobalt based free layer (fig.5 in Ref. [12]). This is because Heusler compounds have higher spin polarization factor (P) than Cobalt. Heusler compounds also possess low saturation magnetization ($M_s$) and low damping constant ($\alpha$) values. Even P is temperature dependent; these compounds will give higher TMR values at low temperatures. Spin polarization factor of CFAS is weakly dependent on temperature.

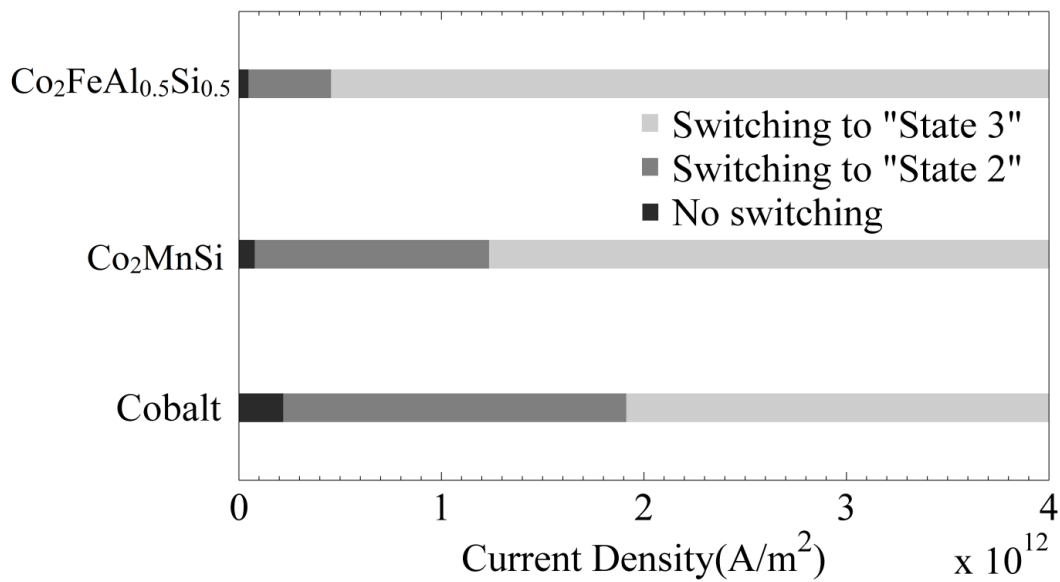

Fig.6. Switching regions to state2 and state3 from state1 by raising the current density values.

## Conclusion:

We investigated the STT switching of cross shaped MTJ by replacing the material of free layer with Heusler compounds. We have shown that switching time and critical switching current density values to switch from one state to other state are less by using heusler compounds as electrodes. We have investigated that for the less value of current density, switching happens faster if we use heusler compounds as electrodes. This could lead to increase in the storage density without increasing the switching current density. Speed of the system to read or write the data also increases as switching time decreases.